\documentclass[preprint,authoryear,12pt]{elsarticle}

\usepackage{graphics}

\usepackage{float}
\usepackage{graphicx}
\usepackage{wrapfig}

\usepackage[draft]{todonotes}   

\usepackage[labelfont=bf]{caption}
\usepackage{caption}
\usepackage{color}
\usepackage{amsmath} 
\usepackage{mathtools} 
\usepackage{amsfonts} 
\usepackage{amssymb}
\usepackage{array,longtable}

\usepackage{array}
\usepackage{booktabs}
\usepackage{rotating}
\usepackage{multirow}
\usepackage{multicol}
\usepackage{lscape}
\usepackage{subfig}

\usepackage[utf8]{inputenc}
\usepackage[T1]{fontenc}
\usepackage{textcomp}
\usepackage{gensymb}

\usepackage[export]{adjustbox}

\usepackage{algorithm,algorithmicx,algpseudocode}

\usepackage{color}

\usepackage{comment}

\usepackage[colorlinks=true,linkcolor=blue,citecolor=blue]{hyperref}

\usepackage[draft]{todonotes}   

\usepackage{soul}

\definecolor{darkgreen}{rgb}{0.53, 0.66, 0.42}
\definecolor{azure}{rgb}{0.0, 0.5, 1.0}

\journal{Medical Image Analysis}

\begin{document}

\begin{frontmatter}



\title{Brain Graph Super-Resolution Using Adversarial Graph Neural Network with Application to Functional Brain Connectivity}

\author[BASIRA]{Megi Isallari}
\author[BASIRA,DUNDEE]{Islem Rekik\corref{cor}}

\cortext[cor]{Corresponding author: irekik@itu.edu.tr; \url{http://basira-lab.com/}.}

\address[BASIRA]{BASIRA lab, Faculty of Computer and Informatics Engineering, Istanbul Technical University, Istanbul, Turkey}
\address[DUNDEE]{School of Science and Engineering, Computing, University of Dundee, UK \ }


\begin{abstract}

Brain image analysis has advanced substantially in recent years with the proliferation of neuroimaging datasets acquired at different resolutions. While research on brain image super-resolution has undergone a rapid development in the recent years, brain \emph{graph super-resolution} is still poorly investigated because of the complex nature of \emph{non-Euclidean graph} data. In this paper, we propose the first-ever deep graph super-resolution (GSR) framework that attempts to automatically generate high-resolution (HR) brain graphs with $N'$ nodes (i.e., anatomical regions of interest (ROIs)) from low-resolution (LR) graphs with N nodes where $N < N'$. First, we formalize our GSR problem as a node feature embedding learning task. Once the HR nodes' embeddings are learned, the pairwise connectivity strength between brain ROIs can be derived through an aggregation rule based on a novel Graph U-Net architecture. While typically the Graph U-Net is a node-focused architecture where graph embedding depends mainly on node attributes, we propose a graph-focused architecture where the node feature embedding is based on the graph topology. Second, inspired by graph spectral theory, we break the symmetry of the U-Net architecture by super-resolving the low-resolution brain graph structure and node content with a GSR layer and two graph convolutional network layers to further learn the node embeddings in the HR graph. Third, to handle the domain shift between the ground-truth and the predicted HR brain graphs, we incorporate adversarial regularization to align their respective distributions. Our proposed AGSR-Net framework outperformed its variants for predicting high-resolution functional brain graphs from low-resolution ones. Our AGSR-Net code is available on GitHub at \url{https://github.com/basiralab/AGSR-Net}.

\end{abstract}


\end{frontmatter}


\section{Introduction}

In recent years, unparalleled neuroscience research studies, namely the Human Connectome Project \citep{hcp}, have optimized and combined state-of-the-art neuroimaging technologies to produce neural data at unprecedently high spatial resolutions. The acquisition and accessibility of these large brain datasets have facilitated methodological studies on systematic mapping of brain connections based on the two most widely used measures of brain connectivity: functional connectivity and structural connectivity, derived from functional magnetic resonance imaging (fMRI) and diffusion weighted imaging (DWI) \citep{marcus}, respectively. Yet, probing trillions of neural connections and axonal pathways is highly susceptible to the resolution of MRI scanners \citep{highres}. To visualize fine-grained variations in brain structure and function or boost the accuracy of brain disorder diagnosis, high spatial resolution of MRI data is of paramount importance. However, the scarcity and high cost of cutting-edge neuroimaging technology that derives data at submillimeter resolutions (e.g. ultra-high field (7 Tesla) MRI scanners) presents a major challenge to the field of neuroscience \citep{trauma}. Specifically, how can we make the acquisition of MRI data at such spatial resolutions possible without resorting to costly neuroimaging modalities? 

Fortunately, the surge of image super-resolution presents formidable opportunities to build predictive methods that learn how to map a brain intensity image of low resolution to an image of higher resolution \citep{bahrami,Chen_2018,ebner}. Recent advances in deep learning have given rise to a plethora of research studies in image super-resolution from the early approaches using Convolutional Neural Networks (CNN) (e.g. SRCNN\citep{dong2014image}) to the state-of-the-art methods such as Generative Adversarial Nets (GAN) (e.g. SRGAN \citep{ledig2016photorealistic}). For instance, \citep{inproceedings} used Convolutional Neural Networks to generate 7T-like MRI images from 3T MRI and recently \citep{qing} used ensemble learning to synergize high-resolution GANs of MRI differentially enlarged with complementary priors. 
However, while intensity images efficiently represent brain connections, modern graph theory also inspired a network representation of a brain connectome which provides powerful insight into the network architecture of the brain.
While a significant number of image super-resolution methods have been proposed for MRI super-resolution, \emph{superresolving brain connectomes (i.e., brain graphs)} remains largely unexplored for several reasons. 

\emph{Firstly}, upsampling (super-resolution) in particular is a notoriously ill-posed problem since the low resolution (LR) connectome can be mapped to a variety of possible solutions in the high resolution (HR) space. \emph{Secondly}, standard image downsampling/upsampling techniques are not easily generalizable to non-Euclidean data due to the complexity of network data. The high computational complexity, low parallelizability and inapplicability of machine learning methods to non-Euclidean data, render image super-resolution algorithms ineffective \citep{cui}.  

To enhance graph embedding learning, recent autoencoder-based models leverage neural networks trained by minimizing the error between the original input and the graph reconstruction performed by the decoder model \citep{kipf2016semisupervised}. However, decoder models focus on graph embedding reconstruction rather than the expansion of graph topology. Our work is focused instead on the increase of the size of the graph while retaining its topological signature, which is an entirely different problem from image super-resolution or graph deconvolution/upsampling. Solutions for graph deconvolution \citep{guo:2018} or graph upsampling (commonly referred to as unpooling in works such as \citep{li:2020,huang:2020}  mainly focus on deconvoluting the node representation back to the original target graph. In other words, these methods are only designed to revert the process of convolution/pooling by either re-distributing the nodes removed in the pooling layer or decoding the high-level features extracted in the convolution layer. Hence the main goal of these techniques is to reconstruct the topological graph information of the input graph while the encoder is trained to minimize this reconstruction error between the original graph and its latent representation. In graph super-resolution, however, the main goal is to increase the number of nodes in the input graph rather than decrease the dimensionality of each node embedding. This increase in the size of the graph is not a re-distribution of previously removed nodes from the original graph: it is a \emph{learned process} that incrementally adds new nodes to a low-resolution graph based on their connectivity strength in order to predict its high-resolution counterpart. \emph{Finally}, to predict a HR connectome from a LR connectome, it is imperative to handle the domain shift (or fracture) between source and target data.

There has been extensive work to overcome the fracture issue between two domains \citep{moreno}. A conventional approach to handle this problem is to integrate a regularization term that enforces the generated HR image to match the distribution of the target HR image. For instance, SRGAN was the first generative adversarial network capable of recovering photo-realistic textures from LR images for $4\times$ upscaling factors \citep{ledig2016photorealistic}. Recently, \citep{anctilrobitaille2020manifoldaware} proposed a cycleGAN-based model that predicts HR diffusion tensor imaging (DTI) from unpaired T1-w MR images. The domain shift was bridged by using adversarial and cycle-consistency losses to minimize the difference between data distributions on a Riemannian manifold of symmetric positive definite $3 \times 3$ matrices. On the other hand, in network representation of connectomes the preservation of realistic connectivity weights allows for a more accurate prediction of HR brain graphs. However, these GAN-based methods were mainly devised for superresolving image data, which might fail in handling geometric medical data such as brain graphs. 

\begin{figure}[!htpb]
\centering
\includegraphics[width=13cm]{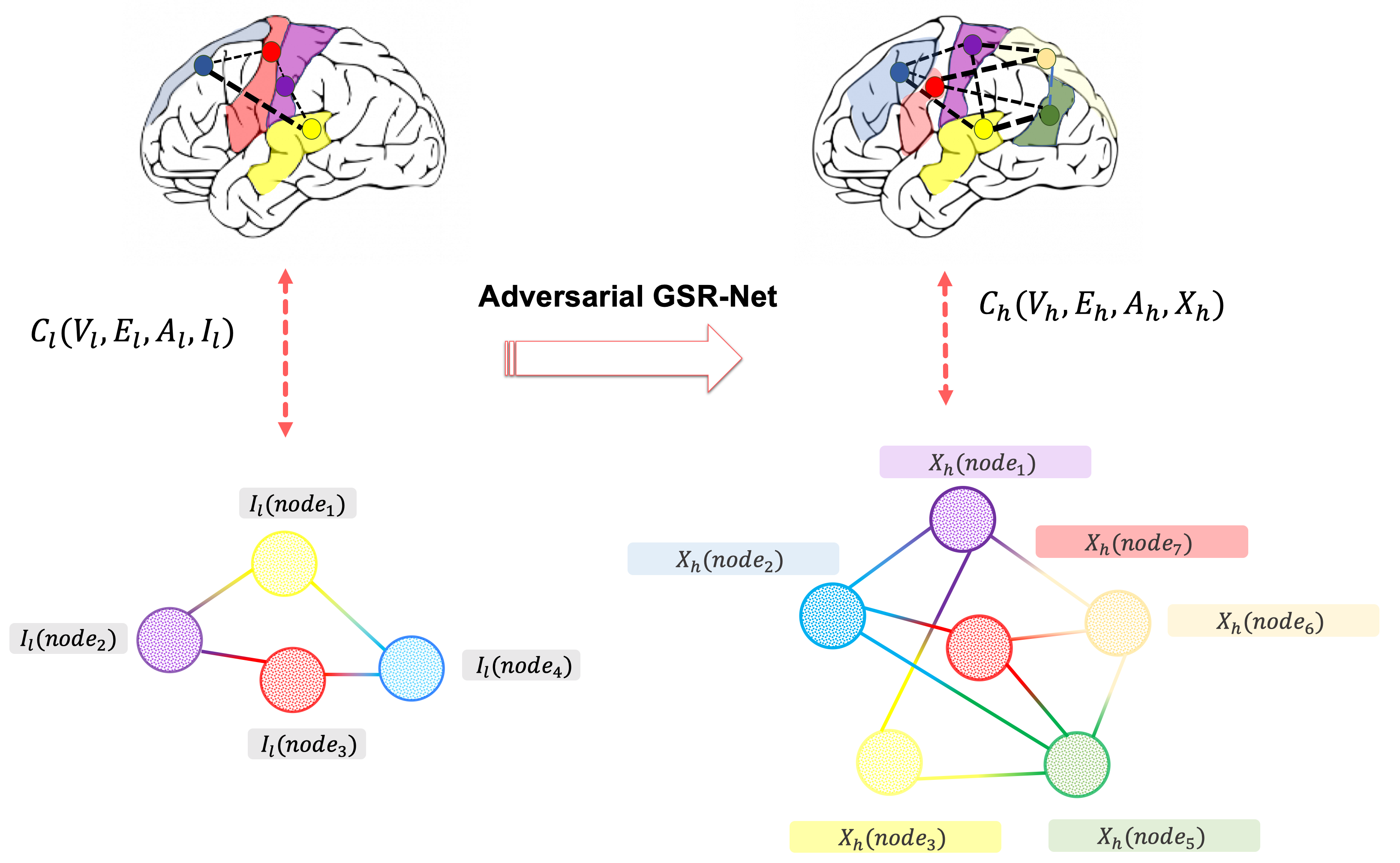}
\caption{\emph{Generation of a high-resolution connectome $\mathbf{C}_{h} = \{ \mathbf{V}_{h}, \mathbf{E}_{h}, \mathbf{X}_{h} \}$ from a low-resolution connectome $\mathbf{C}_{l} =  \{ \mathbf{V}_{l}, \mathbf{E}_{l}, \mathbf{X}_{l} \} $}.  In this illustration, we depict the main purpose of our framework. Each sample in our dataset is represented by a connectome $\mathbf{C}_{l}$ where nodes initially have no features ( $\mathbf{X}_{l} = \mathbf{I}_{l}$ ). Our AGSR learns feature embeddings for each node $\mathit{i}$, encoded in the $i^{th}$ row $\mathbf{X}_l(i)$, in the low-resolution connectome $\mathbf{C}_{l}$ and predicts the topological structure encoded in the adjacency matrix $\mathbf{A}_{h}$ and node embeddings ($\mathbf{X}_{h}$) of the high-resolution connectome $\mathbf{C}_{h}$.}
\label{concept}
\end{figure}

The use of high-resolution brain graphs derived from high-resolution parcellations of brain images was shown to increase the discriminative power of learning-based classifiers in differentiating between different brain states (e.g., healthy versus autistic) \citep{mhiri}. In fact, \citep{mhiri} showed that predicting high-resolution connectomes significantly boosts the diagnosis of autism spectrum disorder (ASD) (i.e. classification of ASD subjects) by 16.48\% in comparison to the original low-resolution connectomes. However, because the low-resolution atlases are most likely to combine interacting ROIs of different temporal signals into a single node,  this adds noise to network analyses and might affect the diagnosis accuracy. In \citep{eickhoff2018imaging} showed the importance of generating brain graphs at different resolutions as these allow to map coarse-to-fine topographies of the brain organization. In fact, using distinctive parcellations of brain images give rise to connectomes with different resolutions, thereby providing a wealth of different features to leverage in computer-aided diagnostic tasks in  network neuroscience \citep{fornito}. Particularly, a brain connectome models the interactions between pre-defined brain parcels following a complex and time-consuming neuroimage pre-processing pipeline \citep{glasser}. As part of the brain connectome generation process, the resolution is simply defined by the parcellation or brain atlas (template) that is used, where the number of parcels (i.e., $N$ anatomical regions of interest (ROIs)) define the connectome resolution (or scale). The resulting parcels form the nodes of the connectome while edges represent measures of structural, functional, or morphological associations between them \citep{fornito,landscape,mahjoub2018,zhu2018,dhifallah:2020}. However, this process has two main drawbacks: (1) initial pre-processing steps are highly prone to variability and bias \citep{qi2015}\citep{bressler} and (2) computational time per subject is very high. Such works and facts have motivated us to address the following problem: \emph{Given a LR connectome, one can devise a systematic method to automatically generate a HR connectome and thus circumvent the need for costly neuroimage processing pipelines.}

To address these limitations, we propose AGSR-Net: the first graph neural network (GNN) framework that attempts to solve the problem of predicting a high-resolution connectome from a low-resolution connectome. AGSR-Net consists of two modules: first, the SRG (super-resolution generator) that learns to generate a HR connectome from the LR connectome and second, the adversarial network that forces the generated HR connectome to match the distribution of the target HR connectome by the discriminator which discriminates whether the current HR connectome comes from the generator or a prior distribution. 
The key idea of AGSR-Net can be summarized in four fundamental steps: (i) learning feature embeddings for each brain ROI (node) in the LR connectome, (ii) the design of a graph super-resolution operation that predicts an HR connectome from the LR connectivity matrix and feature embeddings of the LR connectome computed in (i), (iii) learning node feature embeddings for each node in the super-resolved (HR) graph obtained in (ii), (iv) integrating an adversarial model that acts as a discriminator to distinguish whether a HR connectome is from a prior ground-truth HR distribution or the generated HR connectome in (iii).

\emph{First}, we adopt a U-Net like architecture and introduce the Graph U-Autoencoder. Specifically, we leverage the Graph U-Net proposed in  \citep{unets}: an encoder-decoder architecture based on graph convolution, pooling and unpooling operations that specifically work on non-Euclidean data. However, as most graph embedding methods, the Graph U-Net focuses on typical graph analytic tasks such as link prediction or node classification rather than super-resolution. Particularly, the conventional Graph U-Net is a node-focused architecture where a node $n$ represents a sample and mapping the node $v$ to an $m$-dimensional space (i.e., simpler representation) depends on the node and its attributes \citep{GNNmodel}. Our Graph U-Autoencoder on the other hand, is a graph-focused architecture where a sample is represented by a connectome: a fully connected graph where conventionally, nodes have no features and edge weights denote brain connectivity strength between two nodes. We unify both these concepts by learning a mapping of the node $v$ to an $m$-dimensional space that translates the topological relationships between the nodes in the connectome as node features. Namely, we initially assign identity feature vectors to each brain ROI and we learn node feature embeddings by locally averaging the features of its neighboring nodes based on their connectivity weights. \emph{Second}, we break the symmetry of the U-Net architecture by adding a GSR layer to generate an HR connectome from the node feature embeddings of the LR connectome learned in the Graph U-Autoencoder block. Specifically, in our GSR block, we propose a layer-wise propagation rule for super-resolving low-resolution brain graphs, rooted in spectral graph theory. \emph{Third}, we stack two additional graph convolutional network layers to learn node feature embeddings for each brain ROI in the super-resolved graph. \emph{Finally}, we propose an adversarial regularization of our GSR-Net, where we introduce a discriminator that compares the distribution of the HR graphs generated by the GSR architecture with that of a ground-truth HR distribution. We define our generator as a Graph Convolution Network (GCN) \citep{kipf2016semisupervised} and our discriminator as a multi-layer perceptron.

This work builds on our seed MICCAI Machine Learning in Medical Imaging 2020 workshop paper \citep{isallari2020gsr}\footnote{GitHub: \url{https://github.com/basiralab/GSR-Net}; YouTube video: \url{https://www.youtube.com/watch?v=xwHKRxgMaEM&t=2s&ab_channel=BASIRALab}} by charting the following contributions: (1) Adding a novel adversarial loss for regularization, (2) integrating state-of-the-art benchmark methods including \citep{isallari2020gsr}, (3) using novel topology-driven evaluation metrics to assess the topological soundness of the super-resolved brain graphs and (4) providing an in-depth discussion of the results and the nascent field of graph super-resolution.

\section{Methods}

\subsection{Overall framework}

\textbf{Problem Definition. } A connectome can be represented as an annotated graph \(\mathbf{C}=\{ \mathbf{V},\mathbf{E},\mathbf{A}, \mathbf{X} \}\), where \(\mathbf{V}\) is a set of nodes and \(\mathbf{E}\) is a set of edges connecting pairs of nodes. The network nodes are defined as brain ROIs. The connectivity (adjacency) matrix \(\mathbf{A}\) is an \(N \times N\) matrix (\(N\) is the number of nodes), where $\mathbf{A}_{ij}$  denotes the connectivity weight between two ROIs using a specific metric (e.g., correlation between neural activity or similarity in brain morphology). 
Let \(\mathbf{X} \in \mathbb{R}^{N \times F}\) denote the feature matrix where \(N\) is the number of nodes and $F$  is the number of features (i.e., connectivity weights) per node.  Each training subject \(s\) in our dataset is represented by two connectivity matrices in LR and HR denoted as \(\mathbf{C}_{l} =  \{ \mathbf{V}_{l}, \mathbf{E}_{l}, \mathbf{A}_{l}, \mathbf{X}_{l} \} \) and \(\mathbf{C}_{h} = \{ \mathbf{V}_{h}, \mathbf{E}_{h}, \mathbf{A}_{h}, \mathbf{X}_{h} \}\), respectively. Given a brain graph \(\mathbf{C}_{l}\), our objective is to learn a mapping $f: (\mathbf{A}_{l},\mathbf{X}_{l}) \mapsto (\mathbf{A}_{h},\mathbf{X}_{h})$, which maps $\mathbf{C}_{l}$ onto $\mathbf{C}_{h}$.

\begin{sidewaysfigure}[!htpb]
\centering
\includegraphics[width=20cm]{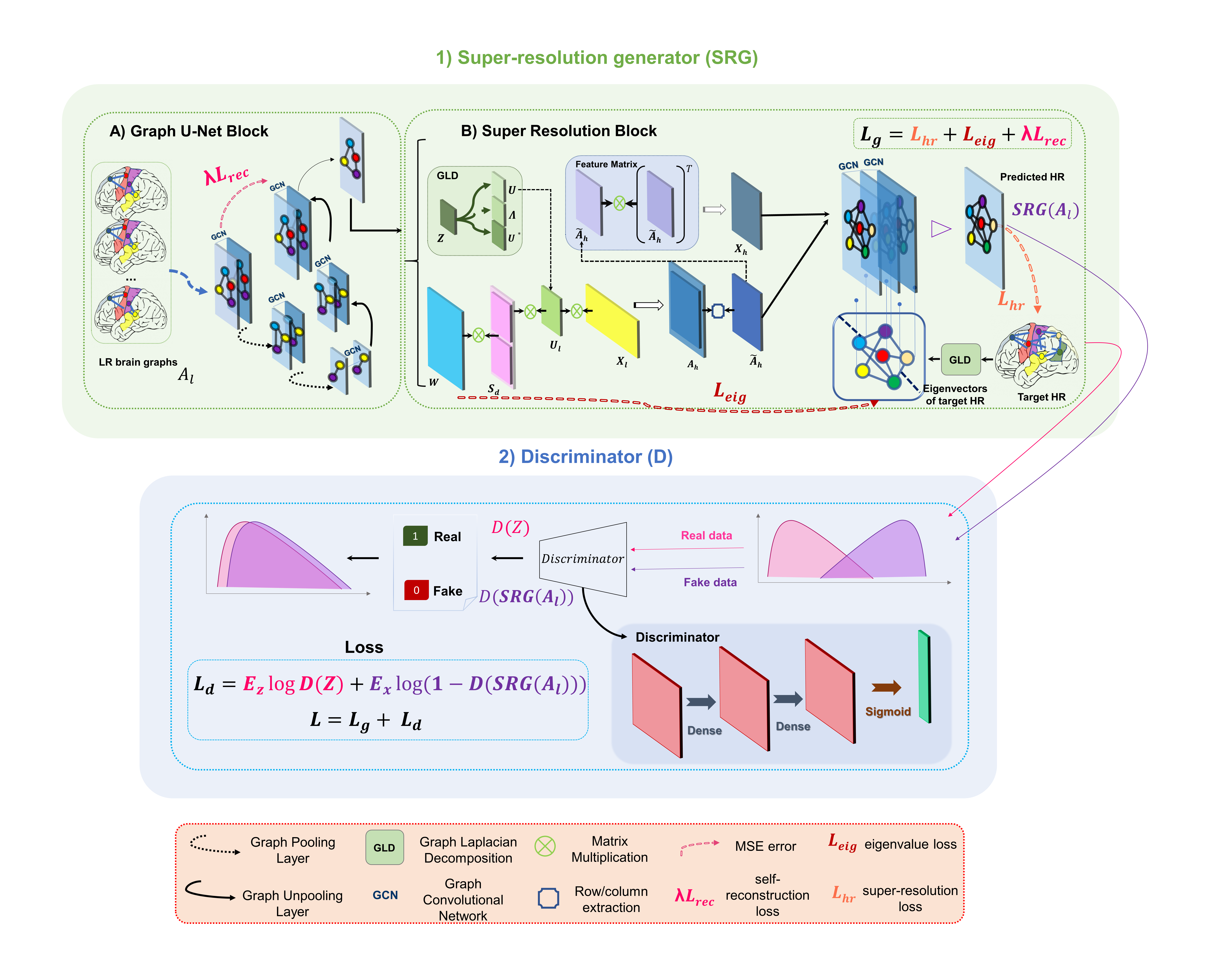}
\caption{\emph{Proposed framework of Adversarial Graph Super-Resolution Network (AGSR-Net) for super-resolving low-resolution brain graphs.} The first three blocks of our architecture comprise the SRG (super-resolution generator) module. \textbf{1) Super-resolution Generator (SRG): } \emph{Graph U-Autoencoder Block.} Our Graph U-Autoencoder is built by stacking two encoding modules and two decoding modules. An encoding module contains a graph pooling layer and a graph convolutional network (GCN) and its inverse operation is a decoding module comprised of a graph unpooling layer and a GCN. Here, we integrate a \emph{self-reconstruction loss} that guides the learning of node feature embeddings for each brain ROI in the LR connectome. \emph{Super Resolution Block.} The GSR Layer super-resolves both the topological structure of the LR connectome (connectivity matrix $\mathbf{A}_{l}$) and the feature matrix of the LR connectome ($\mathbf{X}_{l}$). \emph{Loss function.} The SRG loss comprises a self-reconstruction loss, super-resolution loss and eigen-decomposition loss to optimize learning the predicted HR connectome from a LR connectome. \textbf{2) Discriminator (adversarial regularization).} The adversarial module forces the HR connectome generated by SRG to match a prior ground-truth HR distribution by a discriminator module which identifies whether the connectome instance comes from the prior distribution or the generated HR connectome by SRG.} 
\label{fig1}
\end{sidewaysfigure}

\textbf{Fig.}~\ref{fig1} provides an overview of the key steps of our proposed adversarial graph super-resolution (AGSR) using adversarially regularized graph embeddeding learning and super-resolution rooted in spectral graph theory. Table~\ref{tab1} provides the mathematical notations of all variables used in this paper, specifying their dimensionality and definition. 
\begin{enumerate}
  \item \textbf{Graph U-Autoencoder.} The U-Net takes in the connectivity matrix $\mathbf{A}_{l}$ and node embeddings $\mathbf{X}_{l}$ of LR brain graph where $\mathbf{X}_{l}$ =  $\mathbf{I}_{l}$ and learns the node embedding matrix $\mathbf{Z}_{l}$ with the formal format as follows: $(\mathbf{A}_{l}, \mathbf{X}_{l}) \mapsto \mathbf{Z}_{l}$. Specifically, the U-Net maps a node $\mathit{n}$ to an $\mathit{m}$-dimensional space that translates the topological relationships between the nodes in the connectome as node features. 
  \item  \textbf{Graph Super-Resolution Layer.} GSR Layer takes in the LR brain graph's connectivity matrix $\mathbf{A}_{l}$ and node embeddings $\mathbf{Z}_{l}$ found in (1)  and maps them to a HR connectivity matrix $\mathbf{\tilde{A}}_{h}$ and node embeddings matrix $\mathbf{\tilde{X}}_{h}$  respectively, by learning the transformation $f_{h}: (\mathbf{A}_{l}, \mathbf{Z}_{l}) \mapsto (\mathbf{\tilde{A}}_{h}, \mathbf{\tilde{X}}_{h})$.
  \item \textbf{Stacked GCN layers.} To propagate the topological relationships between the new nodes in our super-resolved graph and learn the HR node embeddings, we stack two graph convolution network (GCN) layers in the form of  $f_{z}: (\mathbf{\tilde{A}}_{h}, \mathbf{\tilde{X}}_{h}) \mapsto \mathbf{Z}_{h}$.
  \item \textbf{Adversarial regularization.} The first three blocks of our architecture comprise the SRG (super-resolution generator) module. The adversarial network forces the HR node embeddings matrix generated by SRG to match a prior ground-truth HR distribution by a discriminator module which distinguishes whether the current matrix comes from a prior distribution or the generated HR node embeddings in (iii). 
\end{enumerate}

\subsection{Super-resolution Generator}

The super-resolution generator ($SRG$) aims to map a low-resolution brain connectome $C(\mathbf{A}_{l}, \mathbf{Z}_{l})$ to the graph embedding of a high-resolution connectome $\mathbf{Z}_{h}$.  

\subsubsection{Graph U-Autoencoder} 

Our Graph U-Autoencoder starts with an initial graph embedding layer (a GCN layer) that aims to learn a coarse node representation of the LR brain graph. This layer takes in the connectivity matrix $\mathbf{A_l}$ and feature matrix $\mathbf{X_l}$ ($\mathbf{X_l} = \mathbf{I_l}$) and generates a node embedding matrix (\(\mathbf{Z}_{0} \in \mathbb{R}^{N \times NK}\)) where $N$ is the number of nodes in the LR connectome and $K$ is the super-resolution factor by which the number of nodes increases in the predicted HR graph with respect to the LR graph ($F$ is specifically chosen to be $NK$ for reasons we explore in greater detail in the next section). 

\begin{table}
\captionsetup{justification=centering}
\caption{Major mathematical notations used in this paper}
\centering
\begin{scriptsize}
\begin{tabular}{ >{\centering\arraybackslash}m{1in} >{\centering\arraybackslash}m{0.5in} >{\centering\arraybackslash}m{3.8in} }
	\toprule
	Notation& 
	Dimension& 
	Definition \\
	\toprule
	$N$ & $\mathbb{N}$ & number of brain regions (i.e, ROIs) \\
	$F$ & $\mathbb{N}$ & number of features extracted from the original brain graph\\
	$E'$ & $\mathbb{N}$ & number of edges in a brain graph \\
	$K$ & $\mathbb{N}$ &  the super-resolution factor by which the number of nodes increases in the predicted HR graph with respect to the LR graph  \\
	$\mathbf{C}=\{ \mathbf{V},\mathbf{E},\mathbf{A}, \mathbf{X} \}$ & $-$ & a brain graph of a specific subject \\
	$\mathbf{C_{l}}=\{ \mathbf{V_l},\mathbf{E_l},\mathbf{A_l}, \mathbf{X_l} \}$ & $-$ & a low-resolution brain graph of a specific subject \\
	$\mathbf{C_{h}}=\{ \mathbf{V_h},\mathbf{E_h},\mathbf{A_h}, \mathbf{X_h} \}$ & $-$ & a high-resolution brain graph of a specific subject \\		
	$\mathbf{V}$ & $\mathbb{N}^{1 \times N}$ & a set of anatomical brain regions representing nodes in the brain graph \\
	$\mathbf{E}$ & $\mathbb{N}^{1 \times E'}$ & a set of edges connecting the brain regions representing either functional, structural or morphological connectivities in the brain \\
	$\mathbf{X}$ & $\mathbb{R}^{N \times F}$ & node embeddings marix of brain graph \\
	$\mathbf{A}$ & $\mathbb{R}^{N \times N}$ & connectivity matrix measuring the pairwise edge weights between nodes (i.e, ROIs)\\
	$\mathbf{A_{ij}}$ & $\mathbb{R}^{1 \times 1}$ & connectivity weight between two ROIs using a specific metric (e.g., correlation between neural activity)\\
	$\mathbf{Z_{0}}$ & $\mathbb{R}^{N \times NK}$ & node embedding matrix of the low-resolution brain graph generated by the initial GCN layer \\
	$\mathbf{u}$ & $\mathbb{R}^{F \times 1}$ & trainable vector in the projection stage of Graph Encoder \\
	$\mathbf{v}$ & $\mathbb{R}^{N \times 1}$ &  vector of scalar projection values of each node on vector u \\
	$indices$ & $\mathbb{R}^{N \times 1}$ &  indices of the $k$ nodes selected for the new downsampled graph \\
	$\mathbf{\tilde{v}}$ & $\mathbb{R}^{k \times 1}$ &  gate vector obtained by applying a sigmoid mapping to each element in the projection vector v \\
	$\mathbf{\tilde{X_l}^{(l)}}$ & $\mathbb{R}^{k \times F}$ & pooled feature matrix according to selected nodes \\
	$\mathbf{X_l^{(l+1)}}$ & $\mathbb{R}^{k \times F}$ & feature matrix of new downsampled graph \\
	$\mathbf{A_l^{(l+1)}}$ & $\mathbb{R}^{k \times F}$ & connecitivity matrix of new downsampled graph \\
	$\mathbf{Z_l}$ & $\mathbb{R}^{N \times NK}$ & node embeddings of low-resolution graph after Graph U-Autoencoder \\
	$\mathcal{L}_{rec}$ & $-$ & self-reconstruction loss: guides the learning of node feature embeddings for each brain ROI in the LR connectome \\
	$\mathbf{L_0}$ & $\mathbb{R}^{N \times N}$ & graph Laplacian of LR brain graph \\
	$\mathbf{L_1}$ & $\mathbb{R}^{N \times NK}$ & graph Laplacian of HR brain graph \\
	$\mathbf{{{U}_{0}}^{*}}$ & $\mathbb{R}^{N \times N}$ & conjugate transpose of the eigenvector matrix of LR brain graph \\
	$\mathbf{U}_{1}$ & $\mathbb{R}^{NK \times NK}$ & eigenvector matrix of HR brain graph \\
	$\mathbf{S}_{d}$ & $\mathbb{R}^{NK \times N}$ & transpose of concatenation of K NxN identity matrices along axis 0 \\
	$\mathbf{W}$ & $\mathbb{R}^{NK \times NK}$ & learnable parameters in GSR layer \\
	$\mathbf{\tilde{X}}_{h}$ & $\mathbb{R}^{NK \times NK}$ & node embeddings matrix of HR brain graph after GSR layer \\
	$\mathbf{\tilde{A}}_{h}$ & $\mathbb{R}^{NK \times NK}$ & connectivity matrix of HR brain graph after GSR layer \\
	$\mathcal{L}_{eig}$ & $-$ & eigendecomposition loss: enforces the eigen-decomposition of the super-resolved connectomes to match that of the ground-truth HR connectomes \\
	$\mathbf{Z_h}$ & $\mathbb{R}^{NK \times NK}$ & node embeddings matrix of HR brain graph after two additional GCN layers \\
	$D(\mathbf{A}_{h})$ & $\mathbb{R}^{1 \times 1}$ & discriminator's estimate of the probability that a target HR brain connectome instance is real \\
	
	$1 - D(SRG(\mathbf{A}_{l},\mathbf{X}_{l}))$ & $\mathbb{R}^{1 \times 1}$ & discriminator's estimate of the probability that an instance of a generated HR connectome by SRG is real \\
	
	$\mathcal{L}_{g}$ & $-$ & super-resolution generator loss: minimizes the error between superresolved brain connectomes and the ground-truth HR ones \\
	$\mathcal{L}_{d}$ & $-$ & discriminator loss: trains the discriminator to distinguish between the generated HR brain connectomes from the ground-truth HR ones \\
	$\mathcal{L}$ & $-$ & overall loss comprised of super-resolution generator loss and discriminator loss \\
\bottomrule
\end{tabular}
\end{scriptsize}
\label{tab1}
\end{table}

The propagation rule of this initial GCN layer can be formulated as follows:
\begin{equation}
\mathbf{Z}_{0} = \sigma(\hat{\mathbf{D}}^{-\frac{1}{2}}\hat{\mathbf{A}}\hat{\mathbf{D}}^{-\frac{1}{2}}\mathbf{X}_l\mathbf{W}_l)
\end{equation}\\\\
where $\hat{\mathbf{D}}$ is the diagonal node degree matrix, $\hat{\mathbf{A}} = \mathbf{A} + \mathbf{I}$ is the adjacency matrix with added self-loops, $\sigma$ is the activation function and $\mathbf{W}_l$ is a matrix of trainable filter parameters to learn. 

To learn node feature embeddings of a given LR connectome \( \mathbf{C}_{l}=\{ \mathbf{V}_{l},\mathbf{E}_{l}, \mathbf{A}_{l} ,\mathbf{X}_{l} \}\),  we propose a Graph U-Autoencoder comprising of a Graph U-Encoder and a Graph U-Decoder. 

\textbf{\emph{Graph U-Encoder}}. In this section, we aim to encode the high-level features for our connectome embedding. The Graph U-Encoder takes in the topological structure $\mathbf{A}_l \in \mathbb{R}^{N \times N}$ of the LR connectome \( \mathbf{C}_{l}=\{ \mathbf{V}_{l},\mathbf{E}_{l} , \mathbf{A}_{l}, \mathbf{X}_{l}\}\) where $N = \mathbf{V}_{l}$ as well as the node content matrix  \(\mathbf{X}_{l} \in \mathbb{R}^{N \times F}\).  To this end, we stack two encoder blocks, each composed of a graph pooling layer and a GCN layer. The graph pooling layer forms a new smaller brain graph by adaptively selecting nodes and the GCN layer aggregates (locally averages) the features of its neighboring nodes based on their connectivity weights.

\emph{Graph Pooling Layer}. Several image-based super-resolution works leverage pooling layers to progressively downsample feature maps and extract the most dominant features in an image. For instance \citep{yu2015multiscale} proposed dilated convolutions to systematically aggregate multi-scale contextual information. \citep{kalchbrenner-etal-2014-convolutional} uses dynamic $k$-Max Pooling, a global pooling operation that selects $k$ maximum values in a linear sequence. However, these operations are based on the spatial locality of pixels in grid-like data which is not inherent among nodes in network representations. To bridge this gap, Graph U-Net proposed a pooling layer specialized in downsampling graph-structured data. Specifically, the pooling layer samples the most significant nodes according to their scalar projection values onto a trainable vector $\mathbf{u}$. 
In this paper, we adopt the graph downsampling concept to decrease the number of brain ROIs by selecting the most significant nodes and thus increasing the local receptive field. However, as a graph-focused approach where the sample is represented by a connectome (i.e. a fully-connected weighted graph), the notion of locality is defined in terms of edge weights rather than node features. The graph's propagation rule can be defined as follows:

\begin{align} 
\mathbf{v} &= \mathbf{X}_l^{(l)} \mathbf{u}^{(l)} /\parallel \mathbf{u}^{(l)}\parallel, \nonumber \\
indices &= rank(\mathbf{v},k),\nonumber \\
\tilde{\mathbf{v}} &= sigmoid(\mathbf{v}(indices)),\\
\tilde{\mathbf{X}_l}^{(l)} &= \mathbf{X}_l^{(l)}(indices,:),\nonumber \\
\mathbf{A}_l^{(l+1)}&= \mathbf{A}_l^{(l)}(indices,indices),\nonumber \\
\mathbf{X}_l^{(l+1)}&=\tilde{\mathbf{X}_l}^{(l)}\odot(\tilde{\mathbf{v}} \mathbf{1}^{T}_F) \nonumber
\end{align}

First, we find the scalar projection of $\mathbf{X}_l$ on $\mathbf{u}$ which computes a one-dimensional v vector, where $\mathbf{v}_i$ is the scalar projection of each node on vector $\mathbf{u}$. Here, to measure the amount of information retained from the original graph during projection, we employ a trainable projection vector $\mathbf{u}$. We adaptively select the subset of nodes with the largest projection values on $\mathbf{u}$ to form a new graph that preserves as much information as possible from the original graph. The rank operation $rank(\mathbf{v},k)$ finds the maximum $k$ values in $\mathbf{v}$ and returns the indices of the nodes selected for the new downsampled graph. We find the feature matrix and connectivity matrix for the new graph by extracting rows and/or columns in the original graph according to the indices of the selected nodes: $\mathbf{X}_l^{(l)}(indices,:)$ and $\mathbf{A}_l^{(l)}(indices,indices)$ respectively.  To control information flow, we compute a gate vector \(\tilde{\mathbf{v}} \in \mathbb{R}^k\) by applying a sigmoid mapping to each element in the projection vector $\mathbf{v}$. We multiply the gate vector \(\tilde{\mathbf{v}} \in \mathbb{R}^k\) with a one-dimensional vector $\mathbf{1}^{T}_F$ of size $F$ where all elements are equal to 1. Finally, we compute the new feature matrix of the downsampled graph $\mathbf{X}_l^{(l+1)} \in \mathbb{R}^{k \times F}$ by performing element-wise multiplication of the product $\tilde{\mathbf{v}} \ \mathbf{1}^{T}_F$ and $\tilde{\mathbf{X}_l}^{(l)}$.\\

\emph{\textbf{Graph U-Decoder}}.
While the Graph U-Encoder's aim was to perform pooling and aggregation operations to encode high-level features, the Graph U-Decoder aims to upsample the new smaller brain graph to its original resolution in order to decode the high-level features. To this end, we design our Graph U-Decoder as the Graph U-Encoder's inverse operation by stacking a decoder module for each encoder counterpart. Each decoder block is composed of a graph unpooling layer to restore the graph's original structure and a GCN layer to aggregate neighborhood information for each node.

\emph{Graph Unpooling Layer}. 
In convolutional neural networks for images, unpooling layers are used to capture object segmentation after the extraction of high-level features. \citep{deconv} proposed a deconvolution network to predict segmentation masks and \citep{lu2019indices} designed a novel indices-based image upsampling operation.
However, these methods are not applicable to network data. We adopt the graph unpooling layer proposed in \citep{unets} to reconstruct the graph's original structure. To this end, we save the indices of the nodes in the new  graph obtained after the pooling operation and we relocate the nodes in their original positions. We can formulate this as follows:
\begin{equation} 
\mathbf{X}^{(l+1)} = relocate(\mathbf{0}_{N \times F},\mathbf{X}_{l}^{(l)},indices)
\end{equation}
where $\mathbf{0}_{N \times F}$ is the reconstructed feature matrix of the new graph (initially the feature matrix is empty) . \(\mathbf{X}_{l}^{(l)} \in \mathbb{R}^{k \times F}\) is the feature matrix of the current downsampled graph and the $relocate$ operation distributes row vectors in $\mathbf{X}_{l}^{(l)}$ into $\mathbf{0}_{N \times F}$ feature matrix according to their corresponding indices stored in $indices$.

\textbf{\emph{Optimization}}. To improve and regularize the training of our graph autoencoder model such that the LR connectome embeddings preserve the topological structure $\mathbf{A}_{l}$ and node content information $\mathbf{X}_{l}$ of the original LR connectome, we enforce the learned LR node feature embedding $\mathbf{Z}_l$ to match the initial node feature embedding of the LR connectome $\mathbf{Z}_{0}$. In our loss function we integrate a \emph{self-reconstruction regularization term} which minimizes the mean squared error (MSE)  between the node representation \(\mathbf{Z}_{0}\) and the output of the Graph U-Autoencoder $\mathbf{Z}_{l}$:
\begin{equation}\label{rec} 
\mathcal{L}_{rec} = \lambda\frac{1}{N}\sum_{n=1}^{N} ||{\mathbf{Z}_{0}}_{i} - {\mathbf{Z}_{l}}_{i}||_2^2.
\end{equation}

\subsubsection{Proposed GSR layer} Recent years have seen remarkable progress in image super-resolution techniques that recover high-resolution images from low-resolution images and videos. While these methods have a wide applicability in medical imaging and brain MRI data specifically, they are not easily generalizable to network representations of the brain (i.e. connectomes). Downsampling and upsampling techniques for graph signals on the other hand, have been recently studied in the field of signal processing \citep{tan1, tan2}. Recently, \citep{Tanaka} proposed a novel upsampling 
technique that upsamples a graph signal while retaining the frequency domain characteristics of the original signal defined in the time/spatial domain. To predict the local connectivity of each node and the global topological structure of the HR graph, we adopt the spectral upsampling concept rooted in graph Laplacian decomposition proposed in \citep{Tanaka}.  
We define upsampling in the graph spectral domain as follows \citep{Tanaka}: 
\begin{equation}\label{signal}
x_{u}[pN + k] = x[k],              p = 0, \dots, C-1
\end{equation}
where \(x\) is the original graph signal, \(x_{u}\) is the upsampled graph signal and the original spectrum is repeated $C$ times.
As a first step, we compute the eigenvector decomposition of the graph Laplacian of the LR brain graph and the HR brain graph, respectively. Let \(\mathbf{L}_{0} \in \mathbb{R}^{N \times N}\) and \(\mathbf{L}_{1} \in \mathbb{R}^{NK \times NK}\) be the graph Laplacians of the LR and HR brain graph, respectively. Mathematically, we write:
\begin{align} 
\mathbf{L}_{0} &= \mathbf{U}_{0} \Lambda {\mathbf{U}_{0}}^{*} \\
\mathbf{L}_{1} &= \mathbf{U}_{1} \Lambda {\mathbf{U}_{1}}^{*}  
\end{align}
where \(\mathbf{U}_{0} \in \mathbb{R}^{N \times N}\) and \(\mathbf{U}_{1} \in \mathbb{R}^{NK \times NK}\).

Our graph upsampling definition in Equation~\ref{signal} can be easily expressed in matrix form as follows:
\begin{equation}
x_{u} = \mathbf{U}_{1}\mathbf{S}_{d}{\mathbf{U}_{0}}^{*}x
\end{equation}
 where $S_d$ is the transposed concatenation of $K$ identity matrices along the horizontal axis: \(\mathbf{S}_{d} = [ \underbrace{\mathbf{I}_{N \times N}   \mathbf{I}_{N \times N} ...}_\text{K}]^T\). We can generalize the upsampling definition in matrix form for a signal $x$ to a signal \(\mathbf{X}_{l} \in \mathbb{R}^{N \times F}\) with \(F\) input channels ($N$ is the number of graph nodes and \(F\) is the dimension of the vector for each node) as follows: 
\begin{equation}\label{a_h}
\tilde{\mathbf{A}}_{h} =  \mathbf{U}_{1}\mathbf{S}_{d}{\mathbf{U}_{0}}^{*}\mathbf{X}_{l}
\end{equation}

Our GSR layer aims to generate a HR graph with $\tilde{\mathbf{A}}_{h} \in \mathbb{R}^{NK \times NK}$. According to equation \ref{a_h}, to generate a \(NK \times NK\) connectivity matrix, the number of input channels $F$ of $\mathbf{X}_{l}$ should be equal to $NK$. Consequently, in our Graph U-Autoencoder we map each node in the LR graph to a $F$-dimensional representation specifically: the output node embeddings matrix \(\mathbf{Z}_{l}\) of the Graph U-Autoencoder is specified to be of the dimensions: \(N \times NK\). $N$ is the number of nodes in the LR connectome and $K$ is the super-resolution factor by which the size (number of nodes) of the predicted HR graph increases with respect to the size of the LR graph.

\emph{\textbf{Super-resolving the graph structure}}. To super-resolve the graph structure of the LR brain graph, we take into account the LR feature matrix learned by the Graph U-Autoencoder and the eigen-decomposition of the LR graph Laplacian. To enforce the eigen-decomposition of the HR graph Laplacian to match that of the target HR graph, we formalize the learnable parameters in this GSR layer as a matrix \(\mathbf{W} \in \mathbb{R}^{NK \times NK}\). The training process will be guided towards the minimization of error between the weights $\mathbf{W}$ and the eigenvectors \(\mathbf{U}_{1}\) of the target high-resolution connectivity matrix \(\mathbf{A}_{h}\). The propagation rule of our layer can be formulated as follows:
\begin{equation}
\tilde{{\mathbf{A}}}_{h} =  \mathbf{W} \mathbf{S}_{d}{\mathbf{U}_{0}}^{*}\mathbf{Z}_{l}
\end{equation} 

\emph{\textbf{Super-resolving the graph node features}}. After expanding the size of the LR graph in our GSR layer, the newly added nodes do not have meaningful representations and our aim is to assign feature vectors to each of the new nodes. Here, we revisit our topology-based feature updating premise to translate topological relationships between all nodes in the super-resolved graph as features. By adding new nodes and edges while attempting to retain the characteristics of the original low-resolution brain graph, it is highly probable that some new nodes and edges will remain isolated, which might cause loss of information in the subsequent layers. To avoid this, we initialize the super-resolved feature matrix \(\tilde{\mathbf{X}}_{h}\) as follows: 
\begin{equation}
\tilde{\mathbf{X}}_{h} = \tilde{{\mathbf{A}}}_{h}   (\mathbf{\tilde{A}}_{h})^T
\end{equation}
This operation links nodes at a maximum two-hop distance and increases connectivity between nodes \citep{chepuri2016subsampling}. Each node is then assigned a feature vector that satisfies this property. Notably, both the adjacency and feature matrix are converted to symmetric matrices to approach more realistic predictions (our target HR connectivity matrix is symmetric):
\begin{align}
 \mathbf{\tilde{A}}_{h} &= (\mathbf{\tilde{A}}_{h} + {\mathbf{\tilde{A}}_{h}}^{T})/2 \\
 \mathbf{\tilde{X}}_{h} &= (\mathbf{\tilde{X}}_{h} + {\mathbf{\tilde{X}}_{h}}^{T})/2
\end{align}

\textbf{\emph{Optimization.}} To learn trainable filters which enforce the eigen-decomposition of the super-resolved connectomes  to match that of the ground-truth HR connectomes (i.e., preserving both local and global topologies), we further add the \emph{eigen-decomposition loss}, which is defined as the the MSE between the weights and the eigenvectors \(\mathbf{U}_{1}\) of the ground-truth high-resolution $\mathbf{A}_{h}$: 

\begin{equation}\label{eig}
\mathcal{L}_{eig} = \frac{1}{N} \sum_{n=1}^{N} ||\mathbf{W}_{i} - {\mathbf{U}_{1}}_{i}||_2^2
\end{equation}

\subsubsection{Additional graph embedding layers} Following the GSR layer, we learn more representative ROI-specific feature embeddings of the super-resolved graph by stacking two additional GCNs: 
\begin{align}
\mathbf{Z}_h^0 &= GCN(\tilde{\mathbf{A}}_h,\tilde{\mathbf{X}}_h) \\
\mathbf{Z}_h &= GCN(\tilde{\mathbf{A}}_h,{\tilde{\mathbf{Z}}_h}^0)\nonumber
\end{align}
For each node, these embedding layers aggregate the feature vectors of its neighboring nodes, thus fully translating the connectivity weights to node features of the new super-resolved graph. The output of this third step constitutes the final prediction of the AGSR-Net of the HR connectome from the input LR connectome. However, our predictions of the HR graph $\mathbf{Z}_h$ are of size $NK \times NK$  and our target HR graph size might not satisfy such multiplicity rule. In such case, we can add isotropic padding of HR adjacency matrix during the training stage and remove the extra-padding in the loss evaluation step and in the final prediction.
\begin{equation}
\mathbf{Z}_h = \mathbf{Z}_h[a: rows - a][a: cols - a]
\end{equation}
where a is $\frac{\arrowvert N_h - \tilde{N}_h \arrowvert}{2}$, $rows$ is the number of rows of $\mathbf{Z}_h$ and cols is the number of columns of $\mathbf{Z}_h$.

\textbf{\emph{Optimization}}. The training process for our SRG is primarily guided by the \emph{super-resolution loss} which minimizes the MSE between our super-resolved brain connectomes and the ground-truth HR ones. The total SRG loss function comprises the \emph{self-reconstruction loss}, the \emph{eigen-decomposition loss}, and the \emph{super-resolution loss} and it is computed as follows: 

\begin{gather}
\mathcal{L}_g =  {\mathcal{L}_{hr}} + {\mathcal{L}_{eig}} + \lambda {\mathcal{L}_{rec}} \\ = 
{\frac{1}{N}\sum_{n=1}^{N} ||{\mathbf{Z}_{h}}_{i} - {\mathbf{A}_{h}}_{i}||_2^2 }  +  {\frac{1}{N}\sum_{n=1}^{N} ||\mathbf{W}_{i} - {\mathbf{U}_{1}}_{i}||_2^2} 
\nonumber \\ +  \lambda{\frac{1}{N}\sum_{n=1}^{N} ||{\mathbf{Z}_{0}}_{i} - {\mathbf{Z}_{l}}_{i}||_2^2}\nonumber
\end{gather}

\subsection{Adversarial regularization of AGSR-Net}

The incorporation of adversarial training has received significant research interest recently. However, few of these approaches are specific to network data (\citep{wang2017graphgan,pan2018adversarially}) and they generally employ adversarial regularization to enhance graph representation learning. In contrast, our approach focuses on introducing regularization to the generated HR brain graph from a LR graph by enforcing the predicted graph to follow the distribution of the ground-truth HR graph.   
In our framework, with the introduction of new nodes and edge connectivities in predicting a HR brain graph from a LR brain graph, it is necessary that the predicted HR data preserves the distribution of the original connectivity weights in our dataset. We propose to adopt a GAN-based adversarial regularization approach to handle the domain shift that might occur between the target HR brain connectomes and the predicted HR brain connectomes from their LR counterparts. 
The adversarial model we use in this paper is motivated by the generative adversarial network (GAN) \citep{goodfellow2014generative}. GAN comprises of two interacting modules: the generator and the discriminator. The key idea of the adversarial model is to enforce the data generated by the generator to match the prior data distribution. To this end, the discriminator distinguishes whether an input sample comes from the prior ground-truth HR distribution or the generator. Simultaneously, the generator is trained to fool the discriminator by generating samples that are indistinguishable from data that comes from the prior distribution. The discriminator model is built on a standard multi-layer perceptron: it comprises of two dense layers and a sigmoid layer that outputs a single value which classifies the input as either real or fake data. ${D(\mathbf{A}_{h})}$ is the discriminator's estimate of the probability that a target HR brain connectome instance $\mathbf{A}_{h}$ is real and ${D(SRG(\mathbf{A}_{l},\mathbf{X}_{l}))}$ is the discriminator's estimate of the probability that an instance of a generated HR connectome by $SRG$ is real. The cost can be computed as follows \citep{goodfellow2014generative}:

\begin{gather*}
\mathcal{L}_d = - \frac{1}{2} \mathbf{E}_{p_{(real)}}[\log{D}(\mathbf{A}_{h})] \\\\
- \frac{1}{2}\mathbf{E}_{p_{(fake)}}[\log{(1 - D(SRG(\mathbf{A}_{l},\mathbf{X}_{l})))}] \nonumber
\end{gather*}

The overall cost of our framework is: 
\begin{equation}
\mathcal{L} = \mathcal{L}_g + \mathcal{L}_d 
\end{equation}
We can formulate the adversarial regularization of the generated HR connectome as the minimization of the cross-entropy cost for training the discriminator as follows: 

\begin{gather}
     \min_{SRG} \max_{D} \mathbf{E}_{p_{(real)}}[\log{D}(\mathbf{A}_{h})] + \\
    \mathbf{E}_{p_{(fake)}}[\log{(1 - D(SRG(\mathbf{A}_{l},\mathbf{X}_{l})))}] \nonumber
\end{gather}

\section{Experiments} 

\subsection{Connectomic dataset and parameter setting} We evaluated our framework on 277 subjects  from the Southwest University Longitudinal Imaging Multimodal (SLIM) study \citep{slimstudy}, which includes brain and behavioral data across a long-term retest-duration within three and a half years, multimodal MRI scans provided a set of diffusion, structural and resting-state functional MRI images, along with rich samples of behavioral assessments addressed --cognitive, demographic and emotional information \citep{dataset}. After several preprocessing steps for resting-state fMRI using Preprocessed Connectomes Project Quality Assessment Protocol, each MRI is parcellated using two different atlases. The first one is the widely used Shen functional brain atlas where, for each subject, 268 regions of interest (ROIs) were generated \citep{SHEN2013403}. The second functional brain atlas is Dosenbach atlas which parcellates the brain into 160 anatomical regions, thereby producing functional brain graphs at $160 \times 160$ resolution \citep{Dosenbach1358}. We used 70\% of our samples as training samples and 30\% as testing samples. \\

We conducted our experiments with a graph U-Autoencoder without skip connections to simplify the training of its deep layers. Since the number of the stacked layers in our graph U-Autoencoder is significantly lower in comparison to the original U-Net architecture, we observed that our model did not overfit and adding skip paths between layers was ineffective. We used a nested grid search to find the optimal combination of the self-reconstruction regularization loss parameter and the learning rate of our optimizer. Our AGSR-Net uses Adam Optimizer with a learning rate of $0.0001$ for both the generator and discriminator models and the number of neurons in both Graph U-Autoencoder and GCN layers is set to $NK$. In our dataset, $N = 160$ and we set $K = 2$ since it is the smallest multiplication factor ($K>1$) that allows to derive a resolution of 268 nodes  from a resolution of 160 nodes. Hence, our predicted HR connectome is of size $NK \times NK = 320 \times 320$. During the training process, to match the dimensions of the predicted HR graph and target HR graph, we added isotropic padding to the target HR connectivity matrix and obtained a $320 \times 320$ matrix from the original $268 \times 268$ matrix. For instance, if we were to super-resolve a 160-resolution connectome up to a 400-resolution connectome, our super-resolution factor $K$ would be 3. Hence, our predictions of the high-resolution connectivity matrix would be of size $480 \times 480$ and to respect the dimensionality, we would add $40 \times 40$ zero padding along both axes of the target high-resolution graph (initially of size $400 \times 400$). During both training loss computation and testing stages, we would unpad both the predicted and target high-resolution matrices and calculate the error between all pairs of the $400 \times 400$ matrices. We believe this does not significantly affect the accuracy of our model because the connectivity strength of these ``artificial'' nodes would be zero and they would have no weight in the aggregation of nodes in their local neighbourhood during topology-based feature updating. However, far from ideal, such a padding strategy might slightly affect the generation of a wholly realistic high-resolution graph and we intend to explore other solutions in the future.\\ 
We set the parameter $\lambda$ of the self-reconstruction regularization loss to $0.1$.

\begin{figure}[!htpb] 
\centering
\subfloat[]{
  \includegraphics[width=8cm]{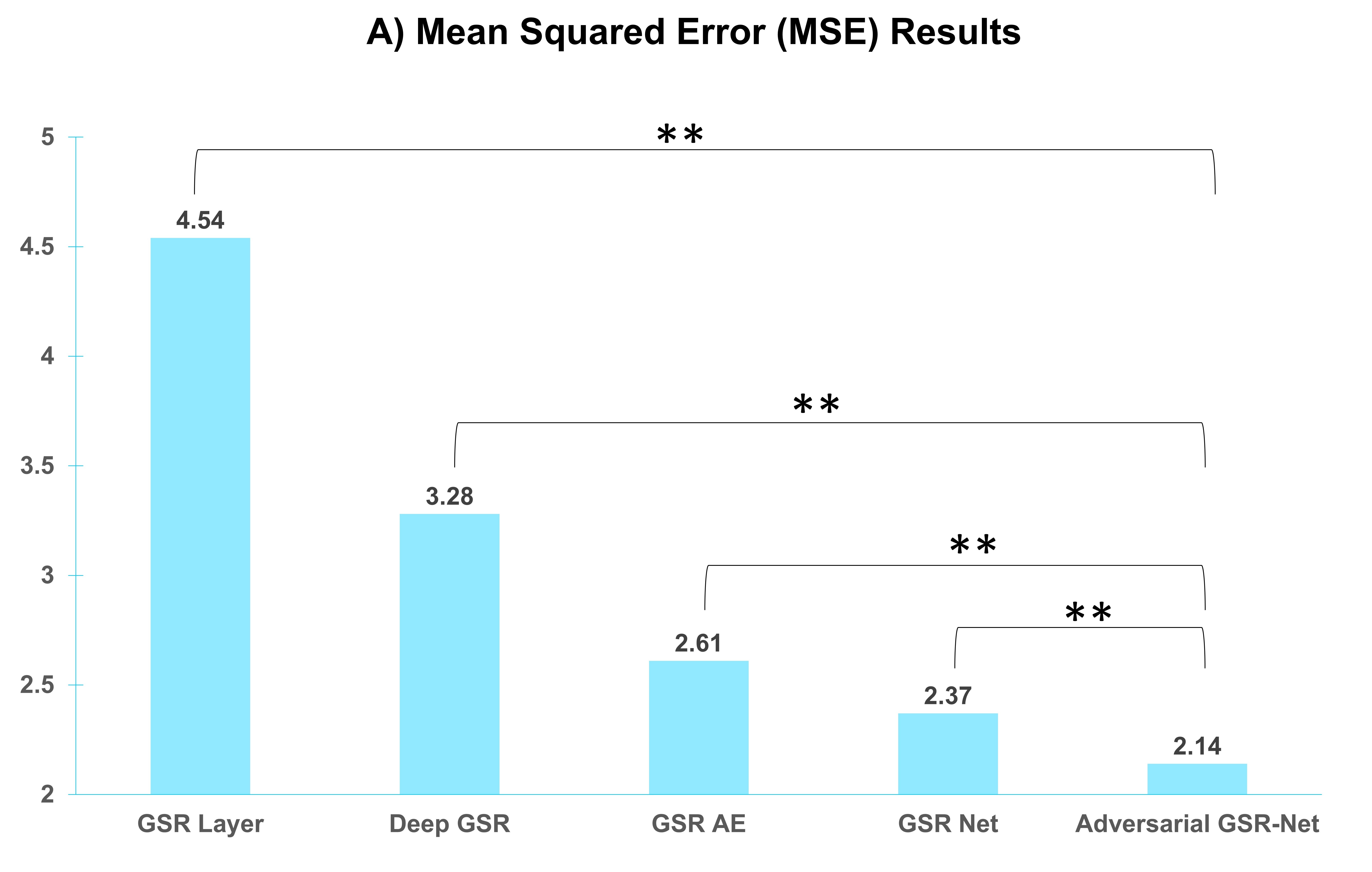}
}
\subfloat[]{
  \includegraphics[width=8cm]{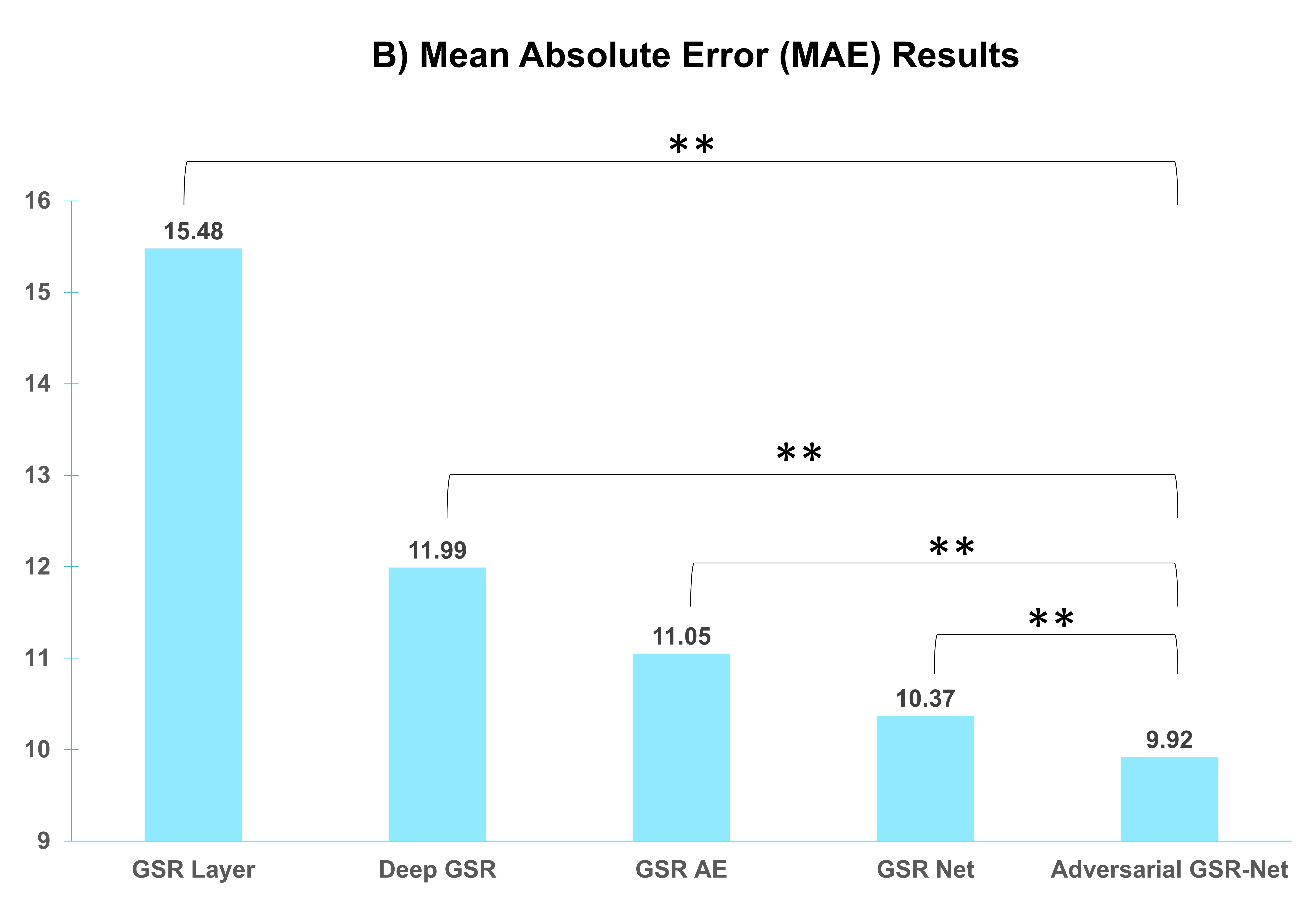}
}
\hspace{0mm}
\subfloat[]{
  \includegraphics[width=8cm]{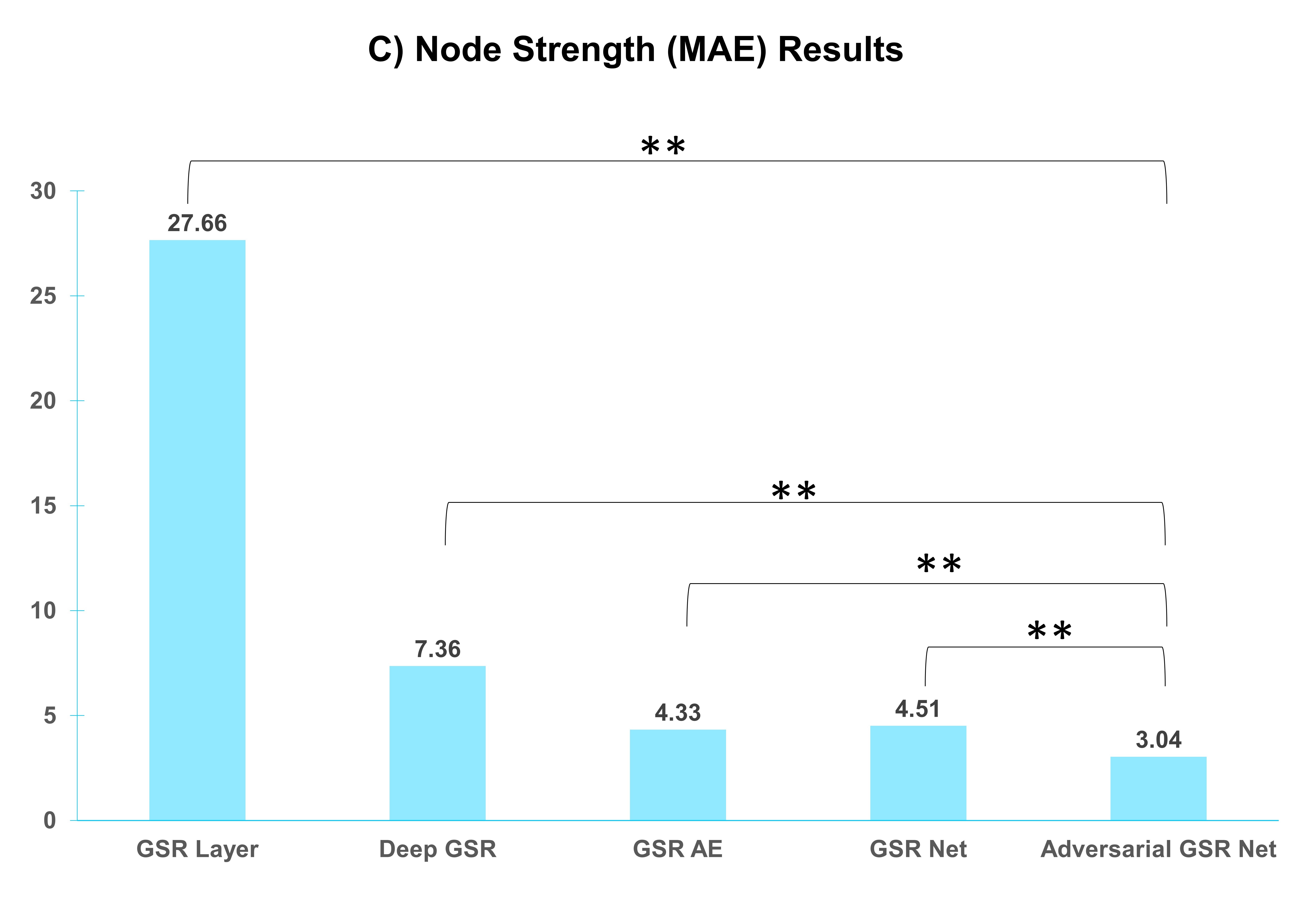}
}
\subfloat[]{
  \includegraphics[width=8cm]{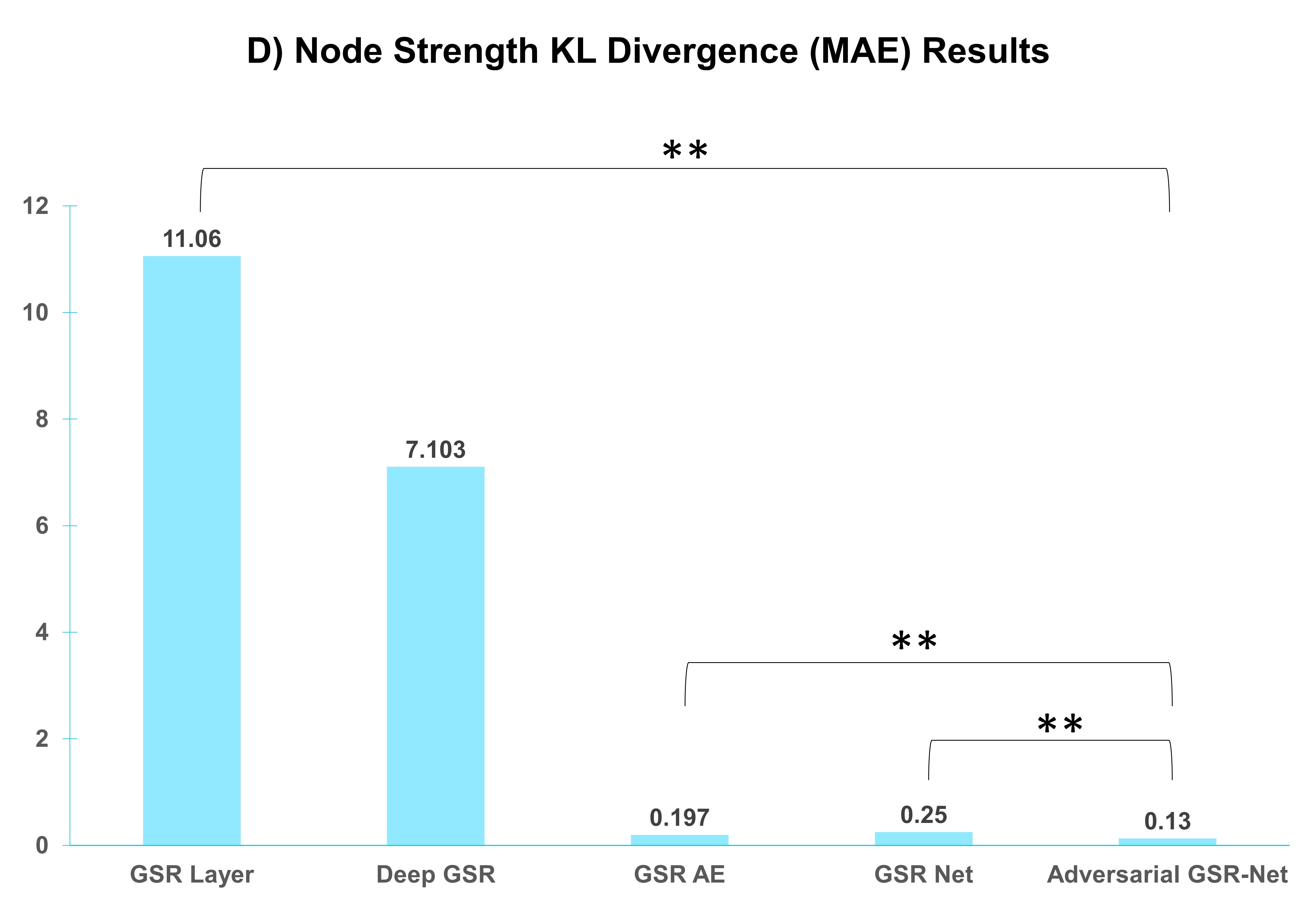}
}
\caption{\textbf{Evaluation of the topological soundness of the generated HR graphs.} We plot the results for for mean squared error (MSE) in \textbf{A}, mean absolute error (MAE) in \textbf {B}, node strength (MAE) in \textbf{C} and node strength KL divergence in \textbf {D} for each of the four baseline methods and our proposed AGSR-Net. (**) : p-value $< 0.05$ using two-tailed paired t-test.}
\label{fig5}
\end{figure}

\begin{figure}[!htpb]
\centering
\includegraphics[width=13cm]{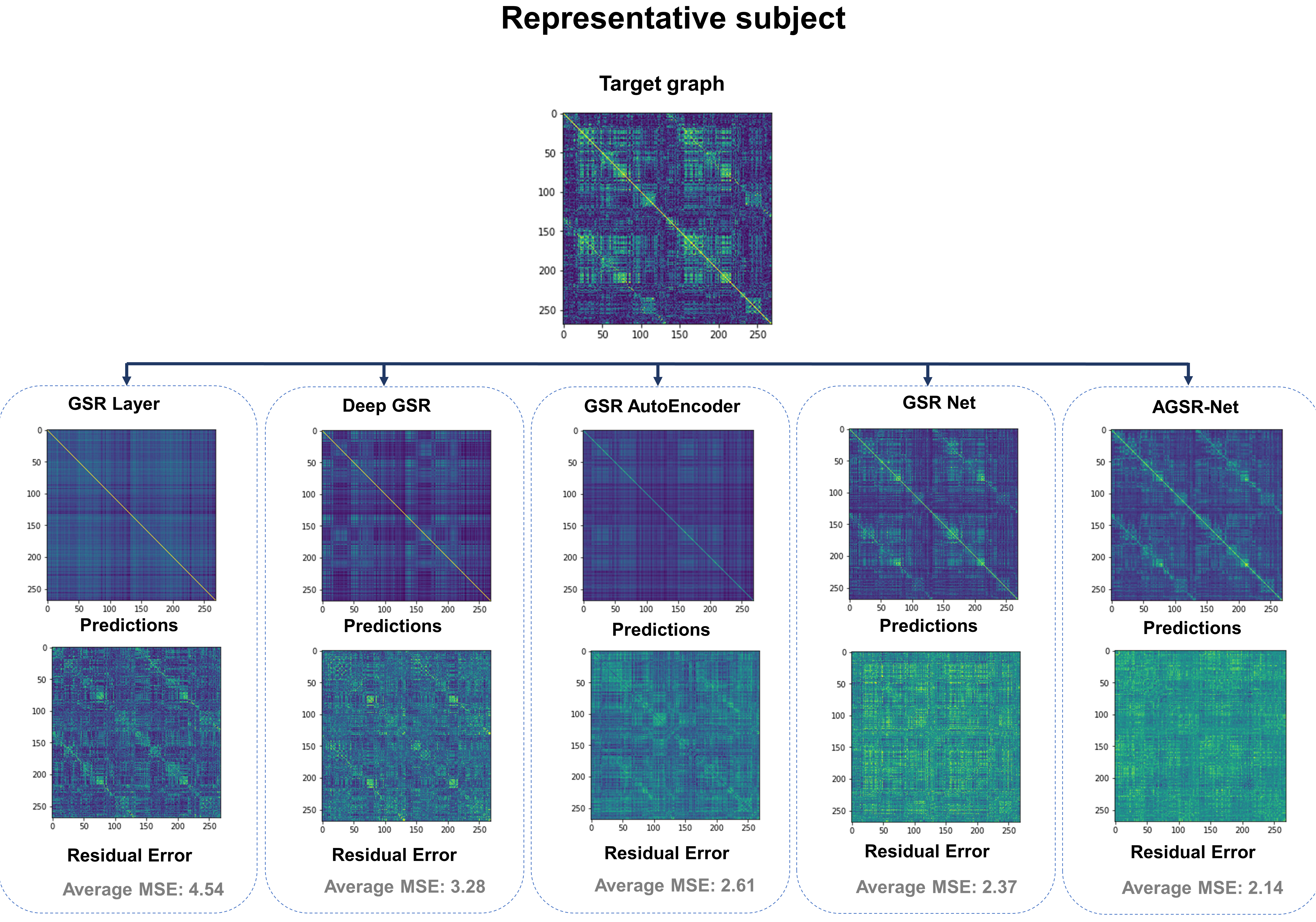}
\caption{\textbf{Comparison between the ground-truth HR graph and the predicted HR graph of a representative subject.} We display the residual error matrix computed using mean squared error (MSE) between the ground-truth and predicted super-resolved brain graph.}
\label{fig6}
\end{figure}

\subsection{Evaluation and comparison methods} We benchmarked the performance of our AGSR-Net against different baseline methods:\\
\textbf{(1) GSR Layer:} a variant of AGSR-Net where we remove both the graph Autoencoder (\textbf{Fig.}~\ref{fig1}--A) and the additional graph embedding layers. In other terms, we directly train the GSR layer for the target graph super-resolution task. This model is primarily guided by the eigen-decomposition loss defined in our paper (Equation ~\ref{eig}). \\
\textbf{(2) Deep GSR:} In this variant, first, the node feature embeddings matrix $\mathbf{Z}_l$ of the LR connectome is learned through two GCN layers. Second, this $\mathbf{Z}_l$ is inputted to the GSR Layer, and third we learn the node feature embeddings  of the output of the GSR Layer (i.e., the super-resolved graph) leveraging two more GCN layers and a final inner product decoder layer. The loss function of this method is modelled after the self-reconstruction loss defined in Equation~\ref{rec}. \\
\textbf{(3) GSR-Autoencoder (AE)}: a variant of AGSR-Net where we remove only the additional GCN layers. \\
\textbf{(4) GSR-Net} \citep{isallari2020gsr}: a variant of AGSR-Net where we remove the adversarial regularization. $\mathcal{L}= \mathcal{L}_g$:  the loss function in GSR-Net is equal to the generator loss of AGSR-Net. 

As illustrated in \textbf{Fig.}~\ref{fig5}, we used the mean absolute error (MAE) and mean squared error (MSE) between the HR brain graphs and the predicted brain graph. Note that the MAE evaluation metric is more robust to outliers. Clearly, our AGSR consistently and significantly (p-value $< 0.05$ using two-tailed paired t-test) outperformed comparison methods in predicting of HR connectomes from LR connectomes.

\textbf{Topology-driven evaluation measures} 
To compare the topological structure of the predicted HR brain graphs with the ground-truth graphs, we further computed (i) the mean absolute error between the node strength of the real and predicted HR graphs and (ii) and the node strength Kullback-Leibler (KL) divergence of the ground-truth HR brain graphs and those of the predicted graphs, respectively. \textbf{Fig.}~\ref{fig5} shows that our method consistently achieves the best topology-preserving predictions compared with its ablated versions. This is an indicator that the nodes with the highest degree (the most significant nodes) in the predicted HR connectome roughly match to the most significant nodes in the target HR connectome.  

\textbf{Fig.}~\ref{fig6} displays the target HR and predicted HR graphs by AGSR and the four comparison methods (GSR Layer, Deep GSR, GSR-Autoencoder, GSR-Net) for two representative subjects. For each subject, we display the target HR brain graph, prediction and the prediction residual. The prediction residual graph is produced by taking the absolute difference between the ground-truth and predicted HR graphs. An average value of the residual matrix is displayed on top of each prediction residual graph. We observe that the residual was significantly ($p < 0.05$) reduced by our AGSR method.

\section{Discussion}

In this paper, we proposed a novel framework to super-resolve non-Euclidean data, specifically brain connectomes or graphs. The main contributions of our model are four-fold:

\noindent \emph{On a conceptual level}, we proposed the first graph super-resolution architecture that aims to automatically generate high-resolution connectomes from low-resolution connectomes using graph neural networks.\\
\emph{On a methodological level}, we proposed to learn node embeddings for the low-resolution brain connectome by translating topology-based relationships as features. Secondly, in our graph super-resolution layer, our contributions were two-fold. Inspired by Tanaka's definition of spectral upsampling for graph signals \citep{Tanaka}, we were the first to introduce a super-resolution layer-wise propagation rule for our graph neural network operating directly on the whole connectome graph structure. Furthermore, we introduced a set of learnable parameters to enforce the eigen-decomposition of the high-resolution graph Laplacian to match that of the target high-resolution graph. Thirdly, we performed adversarial regularization to enforce the distribution of the predicted HR connectome to match that of the target HR connectome. 

In addition, while the theoretical concepts behind our Graph Autoencoder and Graph Super-Resolution layer follow previously published methods, the end-to-end integration of these two concepts to achieve graph super-resolution is not straightforward or simple for the following reasons: \\

\textbf{1)}	Rather than using a general loss function to supervise the training of our model end-to-end, we carefully devised and incorporated loss functions for each of the blocks of our architecture: self-reconstruction loss (ensures that the LR connectome embeddings preserve the topological structure of the original LR connectome) and eigen-decomposition loss (enforces the eigen-decomposition of the super-resolved connectomes to match that of the ground-truth HR connectomes thereby preserving both local and global brain topologies). \\

\textbf{2)}	To prevent loss of information in layers subsequent to our graph super-resolution layer, we increased the connectivity between nodes by multiplying the predicted high-resolution connectivity matrix with its transpose to prevent the potential domain shift caused by the introduction of new nodes and edge connectivities in predicting a HR brain graph. We also employed adversarial regularization to enforce that the predicted HR data preserves the distribution of the original connectivity weights in our dataset. \\
\textbf{3)}	Since our GSR layer super-resolves connectomes by a natural factor, we designed a padding strategy  to account for datasets where the target high resolution is not a multiple of the low-resolution.  We added ``artificial'' nodes via ``graph padding'' to super-resolve any resolution to any resolution —which makes our AGSRNet both generalizable and scalable.\\

\emph{On a computational level}, our method automatically generated a HR connectome which circumvents the need for costly data collection and manual labelling of anatomical brain regions (i.e.parcellation). \\

\emph{On a generic level}, while our framework is tested only on functional MRI data, it can also be applied to connectomes derived from various neuroimaging modalities that represent different node association measures (i.e. structural and morphological connectivity). Also, similar to \citep{mhiri} where the predicted HR connectome significantly boosts autism spectrum disorder diagnosis accuracy, our framework can also as easily integrated into diagnostic network neuroscience pipelines. \\

\textbf{Limitations and future directions.}
Although our proposed method outperformed its variants, there are several limitations that can be addressed in our future work.

\noindent \emph{Computational cost.} In our GSR layer, we compute graph Laplacians for both LR brain graphs and HR brain graphs. To circumvent the high computational cost, in our future work, we will consider well-approximating the eigenvalue vector by a truncated expression in terms of Chebyshev polynomials \citep{hammond2009wavelets}. \\
\emph{Connectome representation.} In this work, all brain connectomes in our data were derived from functional MRI (fMRI). We aim to extend the applicability of our framework to super-resolve connectomes that represent other connectivity measures including structural connectivity derived from diffusion MRI  \citep{Bullmore:2009,honey2009predicting,fornito} and morphological brain networks estimated from conventional T1-weighted MRI \citep{soussia2017,mahjoub2018,nebli2019}. \\
\emph{Multi-resolution brain graphs.} Our method is designed to predict a single high resolution graph from a low-resolution graph. Predicting a \emph{multi-resolution} graph from a source graph (i.e jointly learning brain graphs at multiple resolutions from a single source graph \citep{bressler}) is an intriguing prospect we aim to further explore. \\

\emph{Performance evaluation measures.} In our work, we mainly relied on MSE and MAE to demonstrate the effectiveness of the proposed model. In our future work, we intend to further demonstrate the diagnostic potential of our model using prediction tasks (classification/regression) to show that the generated high-resolution connectome is more informative than the low-resolution connectome and is as informative as the high-resolution connectome calculated using the brain atlas directly. This requires the collection and access of multi-resolution diagnostic datasets. \\

\section{Conclusion}

In this paper, we proposed a novel graph super-resolution framework that attempts to automatically generate high-resolution brain graphs from low-resolution graphs using adversarial graph neural networks. Our framework introduced topology-based node embedding learning of the LR graph, proposed a novel super-resolution layer based on graph Laplacian decomposition and incorporated adversarial regularization of the HR graph. Our method outperformed the baseline methods on our dataset. Future work includes refining our spectral upsampling theory towards fast computation, enhancing the scalability and interpretability of our AGSR-Net architecture with recent advancements in geometric deep learning, and extending its applicability to large-scale multi-resolution brain connectomes.

\section{Acknowledgements}

This project has been funded by the 2232 International Fellowship for Outstanding Researchers Program of TUBITAK (Project No:118C288, \url{https://basira-lab.com/reprime/}). However, all scientific contributions made in this project are owned and approved solely by the authors.

\newpage
\bibliography{references}
\bibliographystyle{model2-names}
\end{document}